\documentclass[mathleft]{an}
\usepackage{graphicx}
\usepackage{times}
\def\src{Cen~X--4}

\def\approxgt{\ifmmode \rlap{$>$}{}_{{}_{{}_{\textstyle\sim}}} \else%
$\rlap{$>$}{}_{{}_{{}_{\textstyle\sim}}}$\fi} 
\def\approxlt{\ifmmode \rlap{$<$}{}_{{}_{{}_{\textstyle\sim}}} \else%
$\rlap{$<$}{}_{{}_{{}_{\textstyle\sim}}}$\fi}

\def\xmm{XMM-{\it Newton}}
\begin{document}

\Pagespan{1}{}
\Yearpublication{2006}%
\Yearsubmission{2005}%
\Month{11}%
\Volume{999}%
\Issue{88}%

\title{Constraining the neutron star equation of state using \xmm}

\author{P.G.~Jonker\inst{1,2,3}\fnmsep\thanks{Corresponding author:
  \email{p.jonker@sron.nl}\newline}
\and  J.~Kaastra\inst{1,3}
\and  M.~M\'{e}ndez\inst{1,4}
\and  J.J.M.~In 't Zand\inst{1,3}
}
\titlerunning{\xmm\, constraints on the EoS}
\authorrunning{Jonker et al.}
\institute{
SRON, Netherlands Institute for Space Research, 3584~CA, Utrecht, The Netherlands
\and 
Harvard--Smithsonian  Center for Astrophysics, Cambridge, MA~02138, Massachusetts, U.S.A.
\and 
Astronomical Institute, Utrecht University, 3508 TA, Utrecht, The Netherlands
\and
Kapteyn Astronomical Institute, Groningen University, 9700 AV, Groningen, The Netherlands
}

\received{12 Oct 2007}
\accepted{12 Oct 2007}
\publonline{later}


\abstract{
We have identified three possible ways in which future \xmm\, observations can provide significant constraints on the equation of state of neutron stars. First, using a long observation of the neutron star X-ray transient Cen~X--4 in
quiescence one can use the RGS spectrum to constrain the interstellar
extinction to the source. This removes this parameter from the X-ray spectral
fitting of the pn and MOS spectra and allows us to investigate whether the
variability observed in the quiescent X--ray spectrum of this source is due to
variations in the soft thermal spectral component or variations in the power law
spectral component coupled with variations in ${\rm N_H}$. This will test whether the
soft thermal spectral component can indeed be due to the hot thermal glow of the
neutron star. Potentially such an observation could also reveal redshifted spectral lines
from the neutron star surface. Second, \xmm\, observations of radius expansion type~I X--ray bursts might reveal redshifted absorption lines from the surface of the neutron star. Third, \xmm\, observations of eclipsing quiescent low--mass X--ray binaries provide the eclipse duration. With this the system inclination can be determined accurately. The inclination determined from the X--ray eclipse duration in quiescence, the rotational velocity of the companion star and the semi--amplitude of the radial velocity curve determined through optical spectroscopy, yield the neutron star mass. }
\maketitle

\section{Introduction}
Low--mass X--ray binaries are binary systems in which a $\approxlt 1 M_{\odot}$ star
transfers matter to a neutron star or a black hole. A large fraction of the
low--mass X--ray binaries is found to be transient -- the so called soft X--ray
transients (here, we will just refer to them as X--ray transients, see
e.g.~Chen, Shrader, \& Livio 1997). Before the arrival of the \xmm\, and {\it
Chandra} satellites only a few neutron star X--ray transients could be studied in
quiescence (e.g.~the system Cen~X--4 and Aql~X--1; Van Paradijs et
al.~1987; McClintock, Horne, \& Remillard 1995; Campana et al.~1997). Using the \xmm\, and {\it Chandra} satellites many more systems have been
studied in quiescence (see e.g.~Rutledge et al.~2001, Wijnands et al.~2001, 2002, Campana
et al.~2003, Jonker et al.~2003, Tomsick et al.~2004, Heinke et al.~2007 to name but
a few references) and Cen~X--4 and Aql~X--1 were studied in much more detail than was possible before (e.g.~Campana et al.~2004, Rutledge et al.~2002). As we will explain below, these observations can have a profound impact on an important area of astrophysics: determining the neutron star equation of state (EoS). This is one of the ultimate goals of the study of neutron stars.

Recent theoretical progress provides the framework for the interpretation of X--ray observations of quiescent low--mass X--ray binaries (e.g.~Brown, Bildsten, \& Rutledge 1998, Colpi et al.~2001, Zavlin, Pavlov, \& Shibanov~1996, Gaensicke, Braje, \& Romani
2002, Heinke et al.~2006). Theoretically, one expects the neutron star to emit X--rays even after accretion has stopped. This emission can be modelled
by a neutron star atmosphere (NSA) model. The NSA models have four free parameters. The neutron star distance, mass, radius and temperature. Due to the large heat capacity the temperature of the neutron star core is set on
timescales of tens of thousands of years (Colpi et al.~2001) and depends on the
equilibrium between the heating and the cooling rate of the neutron star. The
heating rate depends on the total amount of accreted baryons and the
pycnonuclear reactions taking place a few hundred meters deep in the crust (Haensel
\& Zdunik, 1990; Brown, Bildsten, \& Rutledge 1998; Colpi et al.~2001). The time--averaged mass accretion rate in neutron star transients can be derived from
binary evolution models if the orbital period is known (Kraft, Mathews, Greenstein 1962). The pycnonuclear reactions taking place in the neutron star crust
are described in Salpeter \& Van Horn (1969) and  Kitamura
(2000). The balance between the heating and cooling rates sets the
neutron star core temperature (for a 
review of the cooling properties see e.g.~Yakovlev \& Pethick 2004). In theory, a NSA--fit provides means to measure
the mass and radius of the neutron star and thus constrain the equation of state
(EoS) of matter at supranuclear densities. 

In practice, spectra of neutron star transients in quiescence where the flux is
high enough to allow for a spectral study, can indeed be well--fit by a neutron star
atmosphere model (NSA). Sometimes, additional emission is present at energies above
a few keV, often  quantified by a power--law. When systems are selected for which
the distance is well known (e.g.~sources in globular clusters), neutron star masses
and radii can be determined accurately.  Canonical neutron star values were found
(e.g.~Heinke et al.~2003, Webb \& Barret 2007), rendering support for the
interpretation that the soft X-ray spectral component is due to the NSA. Besides
fitting for the neutron star radius and mass in the NSA modelling, there is another
way to derive a constraint on the neutron star mass. Namely, by investigating
the neutron star temperature. The stringent upper limit on the NSA spectral
component to the luminosity of SAX J1808.4--3658 ($\approxlt$10\%; Campana et
al. 2002, Heinke et al. 2007) and the stringent limit on the luminosity and thus
neutron star temperature in 1H~1905+000 (Jonker et al.~2006, 2007) hint at massive
neutron stars in these two transients. The reasoning behind this is the following:
the upper limit on the thermal spectral component implies low core temperatures,
which implies a rapid release  via enhanced
neutrino emission of the energy produced in the crust. This enhanced neutrino emission can only occur if the neutron
star mass is larger than 1.6--1.7 $M_\odot$ (Yakovlev \& Pethick ~2004). Neutron
stars with masses well above $1.4\,M_\odot$ cannot exist for so--called soft
equations of state (EoS), in which matter at high densities is relatively
compressible (e.g., due to a meson condensate or a transition between the hadron and
quark--gluon phases). An important caveat is that the heating accretion history of
the last several tens of thousands of years has to be determined from binary
evolution models.

In this Manuscript we identify three possible ways in which future \xmm\, observations can provide constraints on the neutron star EoS.

\section{Usage of the \xmm\, RGS spectra}
Even though Heinke et al.~(2003) and Webb \& Barret (2007) find values for the neutron star mass and radius consistent with those of canonical neutron stars, it is still not certain whether the soft thermal spectral component is really due to the cooling neutron star. Substantial variations in the quiescent X--ray luminosity of Aql~X--1 and \src\, have been observed (cf.~Campana et al.~1997, Rutledge et al.~2002, Campana et al.~2003). The question is: is this variability caused by variations in the soft spectral component or by
(coupled) variability in ${\rm N_H}$ and the power--law? Currently, the favoured
explanation is that the power--law spectral component varies in accord with ${\rm
N_H}$ (Campana et al.~2003, 2004). In this way the thermal spectral component can be
kept constant. Alternatively, if the soft thermal component varies on short time
scales there is a problem with the interpretation that it arises from the NSA. The
neutron star mass, radius, distance and temperature cannot vary on short timescales.
This said, it turns out that small observed changes in the neutron star effective
temperature can be explained in light of a NSA model if an outburst or a type I
X--ray burst took place between the observations that provide the evidence for
variability. Namely, small temperature changes can be caused by changes in the heat
blanketing layer below the neutron star atmosphere (Brown, Bildsten, \& Chang 2002).
The heat blanketing layer consists of ashes of nuclear burning produced in type I
X--ray bursts and of a layer of H and He that remains after an outburst. The
thickness of the latter layer varies from outburst to outburst. A thicker layer
means a higher heat conductivity which implies a higher observed effective
temperature for a given (unchanged) core temperature. However, the luminosity in the soft band in Cen~X--4 varies on time scales too short to be explained by variations in the thickness of the H/He layer.

With a long \xmm\, observation of \src\, one can determine ${\rm
N_H}$ by measuring the equivalent width of the oxygen edge observable in
the RGS spectrum. In this way ${\rm N_H}$ can be determined independently from the
broadband spectral fit to the pn and MOS spectra, leaving only the temperature
and normalisation of the soft thermal component and the power-law index and normalisation as free parameters to explain the variability.

\begin{figure*}
\includegraphics[width=90mm,height=140mm,angle=-90]{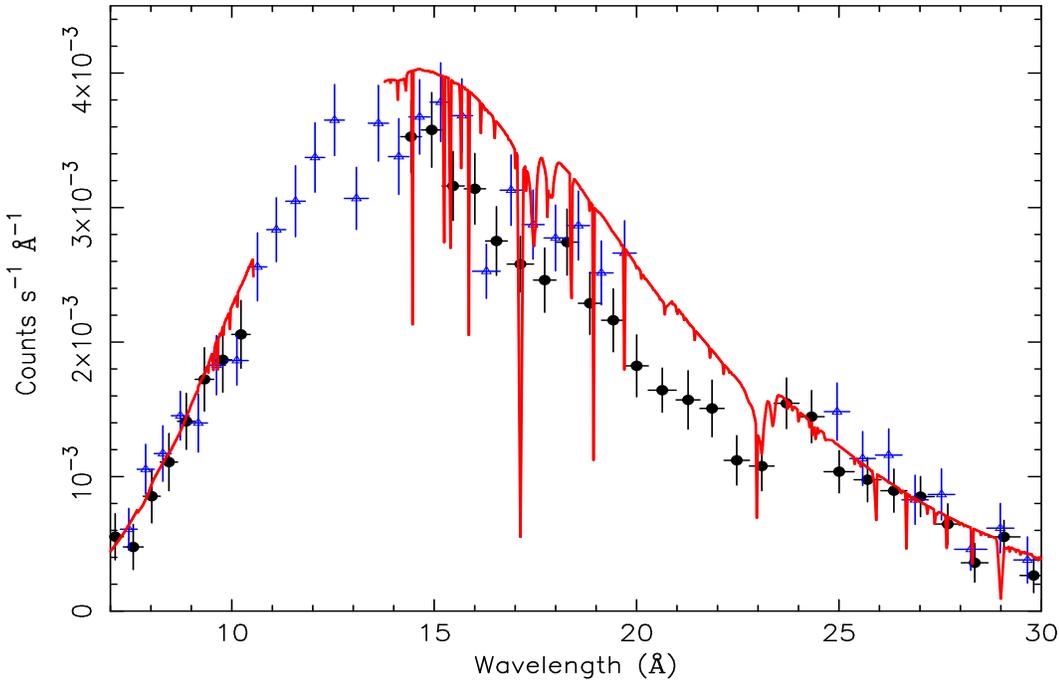}
\caption{A simulated RGS spectrum for a 250 ks--long \xmm\,
observation of \src. The black bullets with error bars are the rebinned simulated
RGS--1 data points, the blue open triangles with error bars are the rebinned
simulated RGS--2 data points, and the red, drawn line is the model fitted to the
RGS--1 data where the O column density is put to zero showing the large difference
between the data and the model especially between 15--22\AA. Most of the narrow
lines are due to CCD gaps and detector drop outs. Unfortunately, most of the O~I
K--edge wavelength is not covered by the RGS--2 due to loss of CCD4 in RGS--2. The
O column and, under the assumption of solar abundances, the ${\rm N_H}$ can
be determined with an accuracy of 8--9\% if the broad band spectral components are
held fixed. Such an observation will 
allow us to determine if indeed the factor of 2--3 variation in  ${\rm N_H}$
necessary to explain the luminosity variations as coupled variations in the power
law index and ${\rm N_H}$ (Campana et al.~2003, 2004) is present or not.}
\label{fig1}
\end{figure*}

There are two possible outcomes of such a study. First, if it is determined that indeed
the soft thermal spectral component does not vary and the variability can be
explained by coupled variations in the power law and ${\rm N_H}$ it would boost the
confidence in the masses and radii that have been and can be determined from fitting
the NSA models to the X-ray spectra of quiescent low--mass X--ray binaries. Note
that unknown uncertainties in the neutron star atmosphere models remain. However, in
this respect it is interesting to note that various atmosphere models give similar
results (see discussion in Webb \& Barret 2007). In the case of \src\, an (HST) parallax distance determination of would allow the most accurate determination the radius of a neutron star to date due to the huge number of photons that will make up the \xmm\, EPIC pn and MOS X--ray spectra. With a distance of 1.2$\pm$0.3 kpc estimated from radius expansion bursts (Chevalier et al.~1989), \src\, is the nearest neutron star X--ray transient currently known.

The second possible outcome is that the soft thermal component varies
substantially. This would either mean that there exists a currently unidentified
mechanism associated with crustal heating that causes the effective temperature to
change on short timescales or that the soft X-ray spectral component is caused by
another process such as residual accretion as proposed by van Paradijs et al.~(1987)
and Zampieri et al.~(1995). This would imply that the cores of these neutron stars
are so cold that the hot thermal glow is not detectable, providing evidence for
enhanced cooling mechanisms in these neutron stars (cf.~Jonker et al.~2007 for the
case of 1H~1905+000).

\section{Redshifted absorption lines from the neutron star surface}

A possible, very important, discovery that might come from deep
observations of quiescent low--mass X--ray binaries is that of redshifted absorption 
lines from the neutron star atmosphere. Currently, redshifted absorption lines might 
have been detected in the bright accreting low--mass X--ray binary EXO 0748--676 (Cottam et al. 2002). 
However, with subsequent observations Cottam et al.~(2007) could not confirm the 
existence of the lines in EXO 0748-676. Redshifted lines have also been searched for 
but not detected in GS 1826-238 (Thompson et al.~2005, Kong et al.~2007).

A potential problem with detecting redshifted absorption lines in quiescent low--mass 
X--ray binaries is that the accretion rate might be too low. Elements more heavy 
than He will sink out of the atmosphere too quickly (Bildsten \& Rutledge 2001). 
Furthermore, for all rapidly rotating neutron stars the contrast between the absorption 
line and the continuum is strongly reduced irrespective of whether the source is in 
quiescence or not. Therefore, one will only under special viewing inclinations be 
able to detect absorption lines in these systems (Chang et al.~2006).

We have identified a special class of type I X--ray bursts that might be better suited 
to search for redshifted metal lines in the atmospheres of neutron stars; radius 
expansion bursts from ultra--compact X--ray binaries. The idea behind grating 
observations of low--mass X--ray binaries showing these type of bursts is that 
the burst luminosity is higher than that of the bursts used for the searches for 
absorption line measurements in EXO 0748-676 and GS 1826-238. Furthermore, 
Weinberg, Bildsten, \& Schatz (2006) showed that the ashes of the nuclear burning 
in radius expansion bursts might contain rare elements that are transported to the 
surface. These elements could produce absorption lines that can be detected in high 
resolution X--ray spectra just after the radius expansion phase when the neutron star 
photosphere settles back on the neutron star.

\section{Determination of the eclipse duration}
Using Kepler's laws and Newtonian Mechanics it has been shown that  the mass of a neutron
star can be determined via the mass function, $f(m)$: $
\frac{M_{NS} \sin^3 i}{(1+q)^2} = \frac{K_{CP}^3\times P}{2 \pi G}$  Here, $M_{NS}$ is the mass of the neutron star, $G$ is the gravitational constant, $P$ the binary orbital period, $i$ the binary system inclination, $K_{CP}$ the radial velocity semi--amplitude of the companion star and $q$ is defined here as the ratio between the mass of the neutron star and that of the companion star. $K_{CP}$ and $P$ can be determined through optical spectroscopic measurements of Doppler shifts of the weak stellar absorption lines as a function of the binary orbital period. To determine the mass
$q$ and $i$ need to be measured as well. 

Tidal interactions between the neutron star and companion star will quickly bring the latter in co--rotation with the orbital motion. This will ensure that the companion star spins rapidly, which manifests itself in a broadening of the stellar absorption lines. In low--mass X--ray binaries broadening of the stellar absorption lines due to effects other  than the rotational velocity of the companion star are at least an order of magnitude smaller. The rotational broadening of the stellar absorption lines can be measured through spectroscopic observations similar to those used to determine $K_{CP}$. One can derive the following relation between the observed stellar rotational velocity ($v \sin i$) and the mass ratio, $q$, in these mass--transferring systems: $\frac{v\sin i}{K_{CP}} = 0.46[(1+q)^2q]^{1/3}$ (e.g.~Wade \& Horne 1988).

The only remaining unknown parameter is the system inclination with respect to our
line--of--sight, $i$. If eclipses in the X--ray lightcurve are observed in these low--mass X--ray binaries simple geometrical arguments show that the inclination has to be $i\approxgt 75^\circ$ (Frank, King, \& Raine 2002). However, knowledge of the mass ratio $q$ and the eclipse duration $\delta \phi$ provides a much more accurate handle on $i$ (Horne 1985).  \xmm\, observations of low--mass X--ray binaries in quiescence can provide the eclipse duration (cf.~Nowak et al.~2002).

\section{Conclusions}

We have identified three possible ways in which \xmm\, can either lead or have an important role in determining the neutron star EoS. First, a long observation of a neutron star low--mass X--ray binary in quiescence with a flux high enough to allow an RGS spectrum to be extracted is necessary to investigate if indeed the soft thermal spectral component is stationary. If so, and if the distance to the system is accurately known the neutron star mass and radius can be determined from a NSA model fit to the EPIC pn and MOS spectra. Currently, the best source for such a study is Cen~X--4. Secondly, one can potentially use radius expansion bursts to search for redshifted absorption lines from the neutron star surface. Again nearby systems are preferred to maximise the source flux. Finally, \xmm\, observations of eclipsing low--mass X--ray binaries yield the eclipse duration, which together with a measurement of the mass ratio yields an accurate determination of the inclination angle that enters the mass function. Using \xmm\, observations mass measurements of these eclipsing sources become much more accurate.

\acknowledgements
PGJ acknowledges support from the Netherlands
Organisation for Scientific Research.



\begin{thebibliography}{}
  \bibitem{} Bildsten, L. \& Rutledge, R.E.:  2001, Ed.~C.~Kouveliotou, J.~Ventura, and E.~van den Heuvel, NATO science series C: Vol 567
  \bibitem{} Brown, E.F., Bildsten, L., Rutledge, R.E.:  1998, ApJ 504, L95
  \bibitem{} Brown, E.F., Bildsten, L., Chang, P.:  2002, ApJ 574, 920
  \bibitem{} Campana, S., Mereghetti, S., Stella, L, Colpi, M.: 1997, A\&A 324, 941 
  \bibitem{} Campana, S., Stella, L.: 2003, ApJ 597, 474
  \bibitem{} Campana, S., Israel, G.L., Stella, L., Gastaldello, F., Mereghetti, S.: 2004, ApJ 601, 474
  \bibitem{} Chang, P., Morsink, S., Bildsten, L., Wasserman, I.: 2006, ApJ 636, L117
  \bibitem{} Chen, W., Shrader, C.R., \& Livio, M.: 1997, ApJ 491, 312 
  \bibitem{} Chevalier, C., Ilovaisky, S.A., van Paradijs, J., Pedersen, H., van der Klis, M.: 1989, A\&A 210, 114
  \bibitem{} Colpi, M., Geppert, U., Page, D., Possenti, A.: 2001, ApJ 548, L175 
  \bibitem{} Cottam, J., Paerels, F., M\'{e}ndez, M.: 2002, Nat 420, 51
  \bibitem{} Cottam, J., Paerels, F., M\'{e}ndez, M., Boirin, L., Lewin, W.H.G., Kuulkers, E., Miller, J.M.: 2007, arXiv:0709.4062
  \bibitem{} Frank, J., King, A., \& Raine, D.J.: 2002, Cam University Press
  \bibitem{} Gaensicke, B.T., Braje, T.M., Romani, R.W.:  2002, A\&A 386, 1001
  \bibitem{} Haensel, P. \& Zdunik, J.L.:  1990, A\&A 227, 431
  \bibitem{} Heinke, C.O., Grindlay, J.E., Lugger, P.M., Cohn, H.N., Edmonds, P.D., Lloyd, D.A., Cool, A.M.: 2003, ApJ 588, 452
  \bibitem{} Heinke, C.O., Rybicki, G.B., Narayan, R., Grindlay, J.E.: 2006, ApJ 644, 1090
  \bibitem{} Heinke, C.O., Jonker, P.G., Wijnands, R., Taam, R.E.: 2007, ApJ 660, 1424
  \bibitem{} Horne, K.: 1985, MNRAS 213, 129
  \bibitem{} Jonker, P.G., M\'{e}ndez, M., Nelemans, G., Wijnands, R., van der Klis, M.: 2003, MNRAS 341, 823
  \bibitem{} Jonker, P.G., Bassa, C.G., Nelemans, G., Juett, A.M., Brown, E.F., Chakrabarty, D.: 2006, MNRAS 368, 1803
  \bibitem{} Jonker, P.G., Steeghs, D., Chakrabarty, D., Juett, A.M.: 2007, ApJ 665, L147 
  \bibitem{} Kaastra, J., de Vries, C.P., Constantini, E., den Herder, J.W.A.: 2007, A\&A in press
  \bibitem{} Kitamura, H.:  2000, ApJ 539, 888
  \bibitem{} Kong, A.K.H., Miller, J.M., M\'{e}ndez, M., et al.: 2007, arXiv:0708.0413
  \bibitem{} Kraft, R.P., Mathews, J., Greenstein, J.L.: 1962, ApJ 136, 312
  \bibitem{} McClintock, J.E.M., Horne, K., Remillard, R.: 1995, ApJ 442, 358
  \bibitem{} Nowak, M.A., Heinz, S., Begelman, M.C.: 2002, ApJ 573, 778
  \bibitem{} Rutledge, R.E., Bildsten, L., Brown, E.F., Pavlov, G.G., Zavlin, V.E.: 2001, ApJ 551, 921
  \bibitem{} Rutledge, R.E., Bildsten, L., Brown, E.F., Pavlov, G.G., Zavlin, V.E.: 2002, ApJ 577, 346
  \bibitem{} Salpeter, E.E. \& van Horn, H.M.:  1969, ApJ 155, 183
  \bibitem{} Thompson, T.W.J., Rothschild, R., Tomsick, J.A., Marshall, H.: 2005, ApJ 634, 1261
  \bibitem{} Tomsick, J.A., Gelino, D.M., Halpern, J.P., Kaaret, P.: 2004, ApJ 610, 933 
  \bibitem{} Van Paradijs, J., Verbunt, F., Shafer, R.A., Arnaud, K.A.: 1987, A\&A 182, 47
  \bibitem{} Wade, R.A. \& Horne, K.: 1988, ApJ 324, 411
  \bibitem{} Webb, N.A., Barret, D.: 2007 arXiv0708.3816
  \bibitem{} Weinberg, N.N., Bildsten, L. \& Schatz, H.: 2006, ApJ 639, 1018
  \bibitem{} Wijnands, R., Miller, J.M., Markwardt, C., Lewin, W.H.G., van der Klis, M.: 2001, ApJ 560, L159
  \bibitem{} Wijnands, R., Heinke, C.O., Grindlay, J.E.: 2002, ApJ 572, 1002
  \bibitem{} Yakovlev, D.G. \& Pethick, C.J.: 2004, ARA\&A 42, 169
  \bibitem{} Zavlin, V.E., Pavlov, G.G., Shibanov, Y.A.:  1996, A\&A 315, 141
  \bibitem{} Zampieri, L.,  Turolla, R., Zane, S., Treves, A.: 1995, ApJ 439, 849
\end{thebibliography}
\end{document}